\documentclass[twocolumn]{aastex631}

\newcommand\new[1]{#1}
\newcommand\old[1]{}

\usepackage{amsmath}

\submitjournal{Astronomical Journal; accepted 16-Mar-2023}

\shorttitle{PSF Regularization}
\shortauthors{Hughes, DeForest \& Seaton}

\graphicspath{{./}{}}

\begin{document}

\title{Coma Off It: Regularizing Variable Point Spread Functions}

\author[0000-0003-3410-7650]{J. Marcus Hughes}
\affiliation{Southwest Research Institute \\
1050 Walnut Street, Suite 300 \\
Boulder, CO 80302, USA}

\author[0000-0002-7164-2786]{Craig E. DeForest}
\affiliation{Southwest Research Institute \\
1050 Walnut Street, Suite 300 \\
Boulder, CO 80302, USA}

\author[0000-0002-0494-2025]{Daniel B. Seaton}
\affiliation{Southwest Research Institute \\
1050 Walnut Street, Suite 300 \\
Boulder, CO 80302, USA}

\begin{abstract}

We describe a \new{rapid and direct} method for regularizing, post-facto, the point-spread function (PSF) of a 
telescope or other imaging
instrument, across its entire field of view.  Imaging instruments in general blur point sources of light
by local convolution with a point-spread function that varies slowly across the field of view, due to
coma, spherical aberration, and similar effects.
It is possible to regularize the PSF in post-processing, producing 
data with a \new{homogeneous} ``effective PSF'' across the entire field of view. In turn, 
the method enables seamless wide-field astronomical mosaics 
at higher resolution than would otherwise be achievable, 
and potentially changes the design trade space for 
telescopes, lenses, and other optical systems where data uniformity is important.  
\new{For many kinds of optical aberration, simple and rapid convolution with a locally optimized ``transfer PSF'' 
produces extremely uniform imaging properties at low computational cost. PSF regularization} does not require access to the instrument that \new{obtained} the data, and can be bootstrapped
from existing data sets that include starfield images \new{or other means of estimating the PSF across the field}.

\end{abstract}


\section{Introduction} \label{sec:intro}

Imaging instruments focus light onto a focal plane.  Focusing is generally not 
perfect: in an ideal instrument, light from each specific direction would be \new{steered}
to an equally specific location on the focal plane; but in real instruments,
even perfectly collimated
incident light is dispersed across a small pattern, the point spread function (PSF), 
that varies slowly with respect to location in the focal plane.  Each region of the 
resulting image is therefore a convolution of the actual data being measured, with the
local PSF for that region of the instrument focal plane.  The PSF is itself a convolution
of image patterns from several effects, including diffraction at the aperture, scattered light, and aberrations caused by the imperfect nature of geometric
optics.  

Reducing optical aberration has been the dominant challenge of telescope optical 
design for some time \citep{KeplerGalileiPena1611,King2003}.  Even in the modern era, 
essentially all imaging instruments, from commercial camera lenses to the Hubble Space Telescope, 
demonstrate aberration of various kinds \citep[e.g.,][]{Smith2008,Khorin_etal2022}, 
including coma, spherical aberration, and position-dependent astigmatism among higher order corrections.
This introduces a slowly-varying PSF that can be quite complicated in shape.  Typical
telescopes may be diffraction-limited near the center of the image plane, but have wide, and/or strongly 
anisotropic, PSFs near the edges of the field of view.  This effect makes telescope resolution a slippery 
quantity: following the Rayleigh criterion of separability of point sources, an instrument may (and most
do) have different spatial resolutions at different focal plane locations or in different directions at the same focal plane location.

Fortunately, the local interaction between a PSF and the underlying image data is well understood
mathematically.  The convolution theorem \citep[e.g.,][]{Bracewell2000} establishes a correspondence
between convolution in the spatial domain and multiplication in the Fourier domain; a corollary is
that the set of all images \new{approximates a commutative group} 
under convolution: just as multiplication by nearly any
number can be reversed through division, the image effects of convolution by nearly any PSF may be 
reversed through \new{convolution by an inverse PSF}, which is essentially division in the Fourier domain.  Using this
insight, one may reverse many PSF-widening effects, improving post facto the effective stray light 
qualities of a telescope through simple transformations of its images; the technique has been used 
to improve the stray light characteristics of solar telescopes in particular,
improving photometry of
dark features embedded in the bright solar corona 
\citep{DeForest_etal2009,Shearer_etal2012}.
Although linear known-PSF deconvolution cannot fully compensate for effects, 
such as diffraction, that literally zero out
particular Fourier components of the image, for some such cases approximate inverses may be constructed that improve
image quality significantly, for example by partially suppressing diffraction lines from
a mesh in front of the instrument \citep{DeForest_etal2009,Poduval_etal2013}.  Further, many common optical aberration 
and stray light ``PSF halo'' effects
do not fully or even deeply suppress the corresponding Fourier components, and therefore can be 
removed through direct known-PSF deconvolution \new{with minimal impact on the image signal-to-noise ratio (SNR)} -- even if \new{the effects} are strongly obtrusive in raw  
images. \new{\citep[e.g.,][]{DeForest_etal2009}}.

\new{In addition to direct deconvolution, many techniques have been developed to transfer images from raw PSF to a desired, more
homogeneous PSF that simplifies or even enables further analysis.  PSF adjustment, with various methods, has been used on a whole-image basis to 
improve slitless imaging spectroscopy \citep{Atwood_2018}, to better match starfield images to models \citep{Alard_1998},  
to improve data comparison across instruments with different PSFs \cite[e.g.,][]{Boucaud_etal_2016}, 
or simply to homogenize atmospheric seeing effects in astronomical surveys \citep[e.g.,][]{Desai_etal_2012}.
}

Because optical aberration generally varies across the field of view, its effects are not susceptible to full-image
linear deconvolution.  However, typical aberration PSF effects such as coma vary ``slowly'', in that the length
scale of variation is much longer than the length scale of the PSF itself.  This means that a sufficiently small region of interest may be deconvolved using linear methods \citep[e.g.,][]{Jarvis_etal2008}.  In the context of 
conventional photography\new{, with a wider region of interest,} variable known-PSF deconvolution has been used for full-frame image 
reconstruction using several commercial compound lenses and even a spherical singlet camera
\citep{Schuler_etal2011}.  The methods described by 
Schuler et al. come at a high computational cost: hours of processing time for
each several-megapixel 
image processed, in addition to the computation required to estimate the PSF
itself.  More recently, \citet{Plowman_etal2022} have developed sparse-matrix
deconvolution methods that offer some advantages and flexibility over those of
Schuler et al. (2011), including both self-consistent treatment of  PSF
variation and also a framework with extensibility to encompass 
more difficult effects such as instrument spectral response or multiple viewing angles and/or constraints \citep{Plowman2022}. 
The computational cost remains high with sparse-matrix methods, because they must represent not \new{merely} a small number of kernels 
but a full PSF realization for every pixel in the original image plane.

\new{
A less rigorous but more practical approach to homogenizing PSF effects across an image 
is to treat the image as a collection of individual patches, each of which may be treated as having 
a separate uniform PSF.  Methods of this sort have been used in a solar physics context to overcome aberration from atmospheric 
seeing \citep[e.g.,][]{Woeger_2007,Scharmer_etal_2010} and, more recently, to address aberration in 
the IRIS solar UV spectrograph \citep{Wuelser_etal_2020}. 
}

In developing the Polarimeter to UNify the Corona and Heliosphere (PUNCH) mission \citep{DeForest_etal2022}, we have
found \new{that} a straightforward method of \new{homogenization}, patched image-neighborhood known-PSF transfer deconvolution, 
can \new{clean} several-megapixel images in a few seconds; this enables routine use of PSF regularization as part of 
\new{the PUNCH data reduction pipeline}, and also as a part of other astronomical or heliophysical image analysis efforts.  
The technique is 
a three-step process: (1) generate a PSF model from bright stars in each subregion of the 
image plane, which may use a single image or a suite of images to ensure enough stars in each 
patch of image; (2) perform analytic known-PSF deconvolution of each image subregion \new{to a target PSF},
using the collection of \new{real} PSFs determined in the first step; (3) merge the image patches
using a soft windowing function, such as the Hann window, to minimize boundary effects in the final
reconstructed image.  This sequence is directly applicable to astronomical imaging; sufficiently 
computationally inexpensive to be applied to survey data or time-dependent data sets, such as image
or spectral-image sequences of the solar corona; and suitable for generating uniform-PSF mosaics from
images acquired by either one camera, as in a sky survey, or many cameras with heterogeneous PSF
characteristics, as for PUNCH.  \new{In particular, simple known-PSF transfer deconvolution is useful for scrubbing PSF
``wings'' and other non-ideal artifacts from images affected by optical aberration, without excessively amplifying 
noise in any part of the images.}

In Section \ref{sec:theory} we briefly introduce the mathematical basis
of known-PSF deconvolution and ``transfer PSFs''; in Section \ref{sec:method} we describe the numerical method and 
sample code to carry out the operation; in Section \ref{sec:demos} we demonstrate de-aberration of model and 
actual starfield data; in Section \ref{sec:discussion} we discuss the implications of routine post facto homogenization
for design of new instruments and analysis of existing data; and in Section \ref{sec:conclusion} we recap and draw
conclusions.

\section{\label{sec:theory}Theory of PSF Deconvolution and Windowing} 

The convolution theorem relates the operations of convolution and
multiplication.  If $I$ and $K$ are images (i.e. $I,K: \mathbb{Z}^2->\mathbb{Q}$ with pixel indices
indicated by subscripts, so that $I_{i,j} \in \mathbb{Q}$ for some range of $i$ and $j$), then
\begin{equation}
    \label{eq:convolution-theorem}
    \left(I\otimes K\right)= \mathcal{F}^{-1}\left[\mathcal{F}\left(I\right)\ \mathcal{F}\left(K\right)\right]\,,
\end{equation}
where $\mathcal{F}$ represents the discrete Fourier transform, $\otimes$ represents convolution, 
and juxtaposition of terms represents elementwise multiplication \new{\citep[e.g.,][]{Bracewell2000}}.  Deconvolution may therefore be 
accomplished through elementwise division in the Fourier plane.  Using $\mathcal{I}$ for $\mathcal{F}(I)$, etc., given a measured $I_K\equiv\left(I\otimes K\right)$, we can immediately write:
\begin{equation}
    \label{eq:deconvolution}
    I = \mathcal{F}^{-1}\left[\frac{\mathcal{I_K}}{\mathcal{K}}\right]\break
       = \mathcal{F}^{-1}\left[\mathcal{I_K}\frac{\mathcal{K}^{*}}{\left|\mathcal{K}\right|^2}\right]\,,
\end{equation}
where: $\mathcal{I_K}=\mathcal{F}\left(I_K\right)$; $\mathcal{K} = \mathcal{F}\left(K\right)$ so that, if K is a point-spread function (PSF), then $\mathcal{K}$ is the corresponding modulation transfer function (MTF); and the $*$ operator represents complex conjugation.  In 
practice, most astronomical images have at least some zero-magnitude Fourier coefficients, so instead of 
the direct reciprocal one \new{may use} a regularized reciprocal\new{\footnote{\new{note the distinction between a ``regularized reciprocal'' which approximates a true reciprocal, and ``PSF regularization,'' which improves and/or homogenizes a slowly-varying image PSF.}}} $\mathfrak{R}$:
\begin{equation}
    \label{eq:recip}
    \mathfrak{R}_{\alpha,\epsilon}\left(\mathcal{A}\right) \equiv \frac{\mathcal{A}^* \left|\mathcal{A}\right|^{\alpha-1}}{\left|\mathcal{A}\right|^{\alpha+1} + \epsilon^{\alpha+1}}\,,
\end{equation}
where $\epsilon$ is a smallness parameter and $\alpha$ controls how sharply the expression drops to zero for small values of $\mathcal{A}$.  Then Equation \ref{eq:deconvolution} becomes
\begin{equation}
    \label{eq:regularized-decon}
    I' = \mathcal{F}^{-1}\left[\  \mathcal{I_K}\  \mathfrak{R}_{\alpha,\epsilon}\left(\mathcal{K}\right)\ \right]\,,
\end{equation}
where the prime on $I'$ indicates that it is a regularized approximation rather than an exact 
reconstruction of the ideal image $I$. \new{Regularized reciprocals are a common tool used to 
approximate the solution of ill-posed inverse problems; 
a good treatment may be found in \citet{claerbout2014geophysical}.}
Equation \ref{eq:regularized-decon} is suitable to deconvolve images that have been convolved with a 
fixed PSF $K$.  The ancillary parameters to $\mathfrak{R_{\alpha,\epsilon}}$ control
the balance between direct analytic inversion of the original PSF, and the need to avoid strongly amplifying noise.  

\begin{figure*}[t]
    \centering
    \includegraphics[width=6.5in]{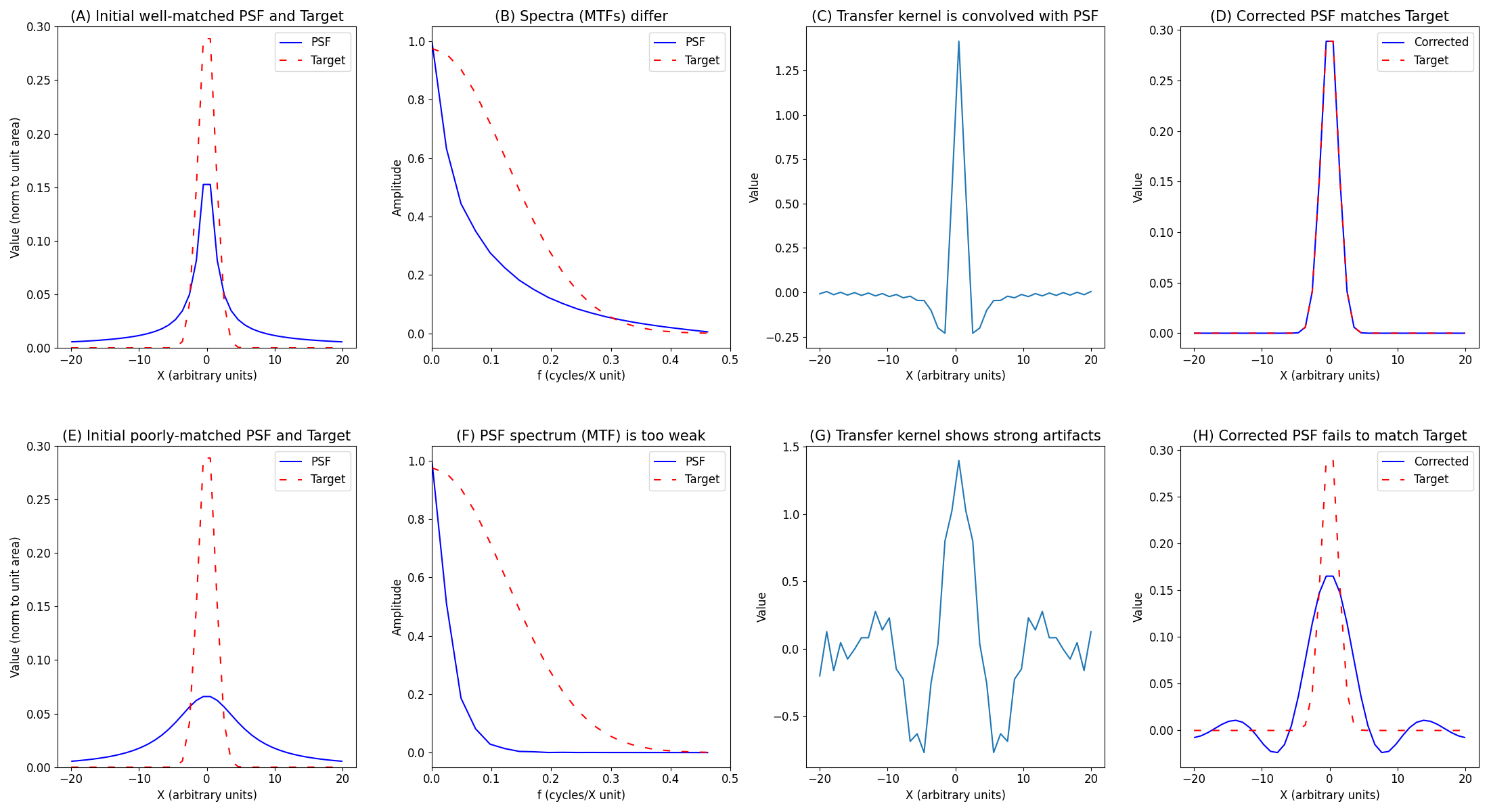}
    \caption{Sample deconvolutions in 1-D demonstrate the effectiveness and limits of PSF \new{correction}.  Top row: a PSF with broad wings (A) moderately attenuates Fourier coefficients at most spatial frequencies, compared to the target PSF (B). Scaling the MTF and Fourier transforming yields a transfer PSF that, when convolved with the original data, matches the target.  Bottom row: a PSF with quite broad wings (E) attenuates the MTF too much for effective correction (F), as seen with the heavily artifacted transfer PSF (G) and poor match between the final effective PSF and the target (H). \new{Note that no noise is added to the data in the figure.}}
    \label{fig:8-panel}
\end{figure*}

Breaking a source image into small neighborhoods allows the inversion in Equation \ref{eq:regularized-decon}
to work for slowly-varying PSFs, such as are caused by optical aberration.  The neighborhoods can be
deconvolved by individually-tailored PSFs $K_i$ for the $i^{th}$ neighborhood.  
However, the optimally achievable degree of deconvolution will in general vary
with the PSF across the image plane.  

To achieve PSF
uniformity, one may convolve the deconvolved images by a band-limited ``target PSF'' $P$, with the property
that $P$ must have zero Fourier coefficients nearly everywhere that each $K_i$ does.  Performing this
operation in the Fourier domain makes clear the importance of band-limiting: zeroed Fourier coefficients
in the band-limited target PSF yield zero coefficients in each deconvolved image:
\begin{equation}
    \label{eq:regularization}
    I_{P,i}' = \mathcal{F}^{-1}\left[\ \mathcal{I}_{\mathcal{K},i}\ \mathfrak{R}_{\alpha,\epsilon,\mathcal{P}}\left(\mathcal{K}_{i}\right)\ \mathcal{P}\ \right]\,,
\end{equation}
where the $i$ subscripts run across neighborhoods of the original image, $\mathcal{P}$ is the desired, constant, band-limited MTF (associated with a target PSF $P$) for all neighborhoods, and $I_{P,i}'$ is
the regularized image neighborhood with the desired uniform PSF applied.  The new quantity
$\mathfrak{R}_{\alpha,\epsilon,P}$ is just:

\begin{equation}    
  \label{eq:recip2}
  \mathfrak{R}_{\alpha,\epsilon,\mathcal{P}}\left(
  \mathcal{A}\right) \equiv 
      \frac{\mathcal{A}^* \left|\mathcal{A}\right|^{\alpha-1}}
           {\left|\mathcal{A}\right|^{\alpha+1} + \left(\epsilon \left|\mathcal{P}\right|\right)^{\alpha+1}}\,,
\end{equation}
\new{in which  $\epsilon$ is multiplied by $\left|\mathcal{P}\right|$ to scale the ``smallness threshold'' to match the band-limited target PSF, limiting overall amplification to roughly $\epsilon^{-1}$.}

The expression $\mathcal{F}^{-1}\left[\ \mathfrak{R}_{\alpha,\epsilon, \mathcal{P}}\left(\mathcal{K}_i\right)\ \mathcal{P}\ \right]$ is interesting, 
because it is convolved with $I_{K,i}$ in Equation \ref{eq:regularization} (implicitly, via Equation \ref{eq:convolution-theorem}), to 
obtain $I_{P,i}'$.  We call this
expression a \textit{transfer PSF} and it can be explicitly convolved with the original image to shift its 
intrinsic PSF from $K_i$ to $P$.

Figure \ref{fig:8-panel} illustrates a correct and incorrect use of transfer PSFs to regularize a 1-D PSF.  In panels (A)-(D), 
data in the form of a delta function have been convolved with a complex curve, based on a Lorentzian, with sharp core and broad wings (bold purple solid line), 
compared to a desired target PSF with slightly broader core and truncated wings (light red dashed line).  Panel (A) shows both this original PSF and a target Gaussian PSF.
Panel (B) shows the modulation transfer function (MTF) of each curve; while the target MTF is higher than the original, the ratio of the two quantities does not exceed roughly 3, for any spatial frequency.  Panel (C) shows a derived transfer PSF, calculated using Equation \ref{eq:regularized-decon}.  Panel (D) shows the result of convolving the original PSF with the transfer PSF in panel (C), matching the target PSF. 

In panels (E)-(H) of Figure \ref{fig:8-panel}, the same process is attempted as in Panels (A)-(D), but with an original PSF whose core is quite broad compared to the target Gaussian curve (and which therefore has less high-spatial-frequency content).  The original MTF is essentially zero for a good fraction of the frequency range in panel (F).  Full transfer 
deconvolution would require amplifying high-frequency Fourier \new{components} by 
large factors, and the calculated transfer kernel (G) has many strong artifacts related to regularization of \new{these components' reciprocals}. In Panel (H), the ``corrected'' PSF does not match the target PSF, because the \new{reciprocal regularization} prevents strong amplification of the missing or near-missing Fourier components (above 0.1 cycles/unit, in Panel F).

In practice, aberration-driven image PSFs retain a sharp core even though the 
wings might be wide or irregular; these PSFs preserve high spatial frequencies,
and therefore afford \new{PSF} regularization to an appropriately-chosen
fixed target PSF that approximately preserves the resolution of the instrument 
as a whole, without greatly amplifying a large set of the corresponding Fourier
components.

\section{\label{sec:method}A Numerical Method for PSF Regularization}

The method described in Section \ref{sec:theory} is implemented and available
as a Python package (with Cython core implementation for speed-up) called 
\textit{regularizePSF} \citep{Hughes_etal_2022a}. Here we define this reference 
algorithm, then discuss the parameters needed for the algorithm and how to 
pick values for them. 

\subsection{\label{sec:algorithm}Description of algorithm}

We perform PSF regularization on a collection of overlapping neighborhoods, 
which are 
recombined at the end of the process to produce a \new{homogenized} image.  The 
neighborhoods are selected to be small enough that the PSF is essentially 
invariant across each one, but large enough that the PSF itself is small 
compared to the neighborhood.  The 
neighborhoods overlap, to ensure smooth splicing in the final recombination.  The algorithm is divided into steps as in the subsections below.

\subsubsection{\label{sec:accumulate-psf}Determine a PSF}

For many astronomical applications, including the PUNCH mission, the scene to be imaged is the dark sky and hence contains many stars.  Because stars are point sources, we use them to measure the PSF in each neighborhood of the instrument field of view.  

We begin with one or more starfield images with the full FOV of the instrument.
We break each image into neighborhoods of size $N\times N$ pixels, with inter-neighborhood spacing of $N/2$ pixels.  This results in sampling each pixel four times into four separate neighborhoods. For each neighborhood, we identify bright stars in the field of view, using a conventional unsharp-mask-and-threshold algorithm, and resample each identified star, together with a small patch of pixels around it, into a common co-aligned image patch of size $M\times M$ pixels (with $M<N$). These identified star image patches are normalized to the same overall flux and then aggregated, using a pixelwise median, into a composite stellar image which serves as a PSF model for the neighborhood.  We use the pixelwise median, rather than the pixelwise mean, to eliminate nearby stars in the small patches around any of the bright stars.  This technique makes use of an implicit assumption that less than half of each star's nearby pixels are affected by asterisms or nebulosity.  Distributed ``objects'' such as nebulosity, zodiacal light, the galaxy or instrumental stray light may be a concern; using a smaller percentile sample than the median can help remove these sources also, or they can be removed via smooth-background estimators such as \textit{minsmooth} \citep{DeForest_etal2016}. Stars near the edge of each neighborhood are sampled in their entirety, even if some peripheral pixels of the corresponding patch are outside the neighborhood boundary; this eliminates boundary effects that otherwise might be associated with each neighborhood.  

Optionally, the accumulated PSF images can be further refined using a parametric fit.  Fitting greatly reduces the residual noise in each PSF image, at the cost of forcing the model PSF to conform to a priori expectations of the form of the fitting function.   Alternatively, one could fit a functional model to either the centered star patches or the averaged stars. Using a functional model may decrease noise in the modeled PSFs and allow for interpolating to fill in regions with insufficient star patches to extract a viable data driven PSF model. However, it requires determining the correct functional form of the PSF and carefully fitting. Parametric fitting is therefore deprecated when noise levels and star counts do not preclude direct aggregation.  Should stellar counts be too low, multiple images from the same instrument may be sampled in each neighborhood, especially with slightly varying pointing, to increase the population of stellar images in each neighborhod.  In applications where stars are not available, other PSF estimation methods, of which there are many \citep[e.g.,][]{Chalmond1991,Starck2002}, may be used instead of direct stellar aggregation or fitting. 

\subsubsection{\label{sec:define-target-psf}Define a target PSF}

We define a target PSF analytically and then evaluate it to create a same-sized patch to the measured (or modeled) local PSF patches. In the examples in Section \ref{sec:demos} we utilize a symmetric 2D Gaussian as the target PSF model. 

The target PSF must be comparable in width to the sharpest core of the measured PSFs.  To design a target PSF, one must inspect the ratio of the two corresponding MTFs, $\mathcal{P}/\mathcal{K}_i$.  If the ratio is large at any location in the Fourier plane, then $P$ is not a suitable target PSF.  The goal of target PSF design is to balance the specific application's need for PSF narrowness, against the need to minimize high amplification factors in any one Fourier component.

\subsubsection{\label{sec:apodize}Apodize each neighborhood}

Before deconvolving each neighborhood, we apodize it by multiplying it with a root-Hann window:
\begin{equation}    
  \label{eq:apod-window}
  w(x, y) = \sin\left( (x+0.5)\frac{\pi}{n_x}\right) \sin\left((y+0.5) \frac{\pi}{n_y}\right)
\end{equation}
where $x$ and $y$ are the current pixel coordinate in the patch and $n_x$ and $n_y$ are the size in pixels of the patches in each dimension. 

The root-Hann window is applied twice: once before, and once after, 
deconvolution.  This minimizes edge effects for the Fourier transform
used in the deconvolution, and also produces Hann window profiles in the 
final \new{homogenized} neighborhoods, simplifying recombination of the neighborhoods (section \ref{sec:merge-regions}).

\new{Apodization is important, in particular, because most Fourier transformation algorithms impose periodic boundary 
conditions on the data being transformed. Combining Hann windows\footnote{\new{Note that many authors have mistakenly referred to the Hann window \citep{Kahlig1993} as a ``Hanning'' window -- a common eggcorn of the related, but different, ``Hamming'' window \citep[e.g.,][]{podder_etal2014}.}} with 50\% overlap is a straightforward 
and common way to segment continuous data, while minimizing spectral spreading from the apodization window itself, 
avoiding boundary effects from the deconvolution step (section \ref{sec:deconvolve}), and simplifying the recombination
of the segments into a continuous whole after processing \citep[e.g.,][]{ISO11172-3,Schlien1994,DeForest2017}. While many Fourier methods ignore apodization entirely \citep[e.g.,][]{ISO10918-1,Handy_etal1999,Wuelser_etal_2020}, such methods are known to introduce neigborhood-boundary artifacts which are detectable in the resulting images whether or not they are directly visible to a user \citep{Schrijver_etal1999,bianchi2011analysis}.
}

\subsubsection{\label{sec:deconvolve}Deconvolve the neighborhoods}

Using the local PSF models and the target PSF for each neighborhood, we deconvolve as in Section \ref{sec:theory}.  In particular, we Fourier transform 
the apodized neighborhood, multiply by the MTF ratio as in Equation \ref{eq:regularization}, then inverse Fourier transform to produce regularized neighborhoods.

\subsubsection{\label{sec:apodize-again}Apply a second apodization}

Because multiplication in Fourier space is equivalent to convolution, the
orginal apodizing window is deconvolved along with the actual image.  This 
produces an incorrect windowing function in the modified neighborhood, as 
described by \citet{DeForest2017}.  To minimize this effect we 
apodize a second time with the root-Hann window as in Section \ref{sec:apodize}.
This yields an overall Hann-window envelope on each neighborhood, subject to 
small perturbations from the deconvolution operation.

\subsubsection{\label{sec:merge-regions}Merge the neighborhoods}

Neighborhoods are re-merged by direct addition, with the appropriate pixel phase, back into a new image array that begins with zero-valued pixels.  
Using the Hann window automatically weights each neighborhood 
to reconstitute the original image without further weighting or scaling to remove the windowing function\new{, because of the identity that $\sin(x)^2 + \cos(x)^2 = 1$.}

\subsection{\label{sec:trades}Algorithm parameters}

The basic \new{PSF} regularization algorithm has several key parameters\new{, described in the paragraphs below}.

The particular target PSF is adjustable.  Ideally the
target PSF should be well localized to preserve resolution, with a core that
is no narrower than the core of the widest neighborhood-PSF core in the source
image plane.  A narrow, isotropic, normalized Gaussian is a good choice.  
In the real-world
examples we considered (section \ref{sec:demos}), 3-pixel FWHM proved to be 
a reasonable minimum width.

The size of each neighborhood must be selected based on the source image
properties.  The size should be small 
enough that the PSF does not change appreciably within each neighborhood, but
large enough that edge effects from the PSF itself do not dominate the
deconvolution operation.  We found that 1/8 to 1/16 the original image size 
was a suitable range in each of our example images.

The $\epsilon$ and $\alpha$ parameters of the $\mathfrak{R}_{\epsilon,\alpha}$ 
operator set the ``smallness'' criterion and the hardness of the transition from 
reciprocal to zero.  The maximum amplification of each Fourier component is
roughly (though in general somewhat less than) $1/\epsilon$, so a value of 
$\epsilon=0.1$ allows Fourier components to be amplified by roughly $10\times$.
$\epsilon$ should be set based on the original signal-to-noise ratio (SNR) of
the source image, and the acceptable SNR in the final image.  In our examples,
we maintained $\epsilon=0.1$.  The $\alpha$ parameter sets the hardness of the
transition from amplification to attenuation of weak components.  We used 
$\alpha=10$; with $\epsilon=0.1$, this allowed up 
to approximately $9\times$ amplification of each Fourier component if needed.

\section{\label{sec:demos}Sample Applications of PSF Regularization}

We applied PSF regularization to a variety of images to demonstrate its effect. 

\subsection{\label{sec:model} Model starfield data}

We first \new{homogenized} a simple model starfield to demonstrate the method in a noiseless setting under controlled conditions.

\begin{figure*}[h]
    \centering
    \includegraphics[width=6in]{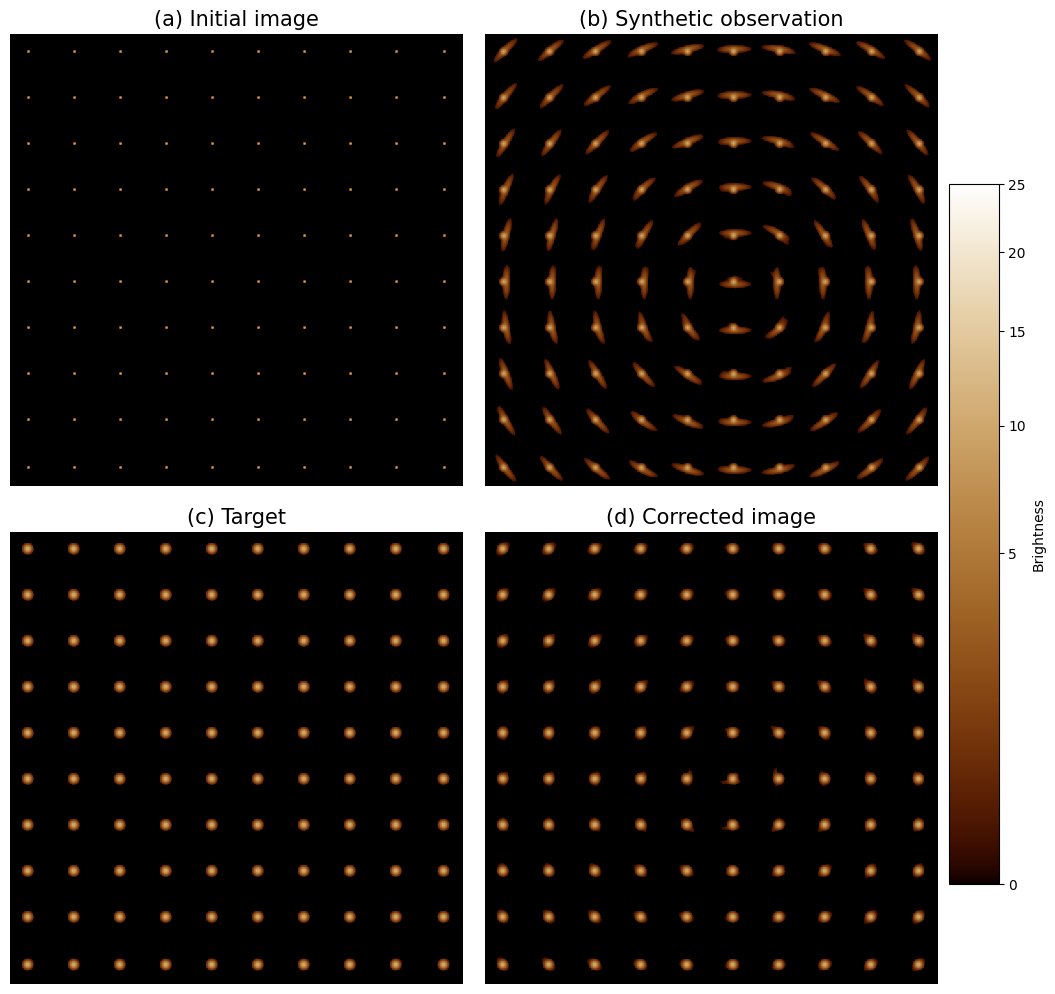}
    \caption{Model data demonstrate PSF \new{homogenization}.  An initial image of a simplified starfield (a) is modified with a slowly varying PSF (b), then \new{homogenized} to a target PSF \new{(c)} with the method of Section \ref{sec:algorithm}.  \new{The final image (d) visually matches the target in (c)}.  The panels are gamma-corrected to highlight the periphery of the model PSFs.
    }
    \label{fig:model}
\end{figure*}

First, we generated point source ``stars'' in a regular square grid across a $256\times256$ pixel image. Next, we separately convolved each star with an individual synthetic instrument PSF calculated from a parametric analytic expression, to create a synthetic observation (Figure \ref{fig:model}, panels (a) and (b)). We applied the algorithm of Section \ref{sec:algorithm} to the image, with a 3 pixel FWHM Gaussian target PSF, resulting in a \new{homogenized} corrected image that matches the convolution of the initial image in Panel (A) with the target PSF (\ref{fig:model}, panels (c) and (d)).  This demonstrates reconstruction of a uniform-PSF image of moderate resolution (panel (d)) from a modeled variable-PSF image (panel (b)).

The details of the synthetic PSF are described in the Appendix. 

\subsection{\label{sec:dash} DASH}

\begin{figure*}
    \centering
    \includegraphics[width=6.5in]{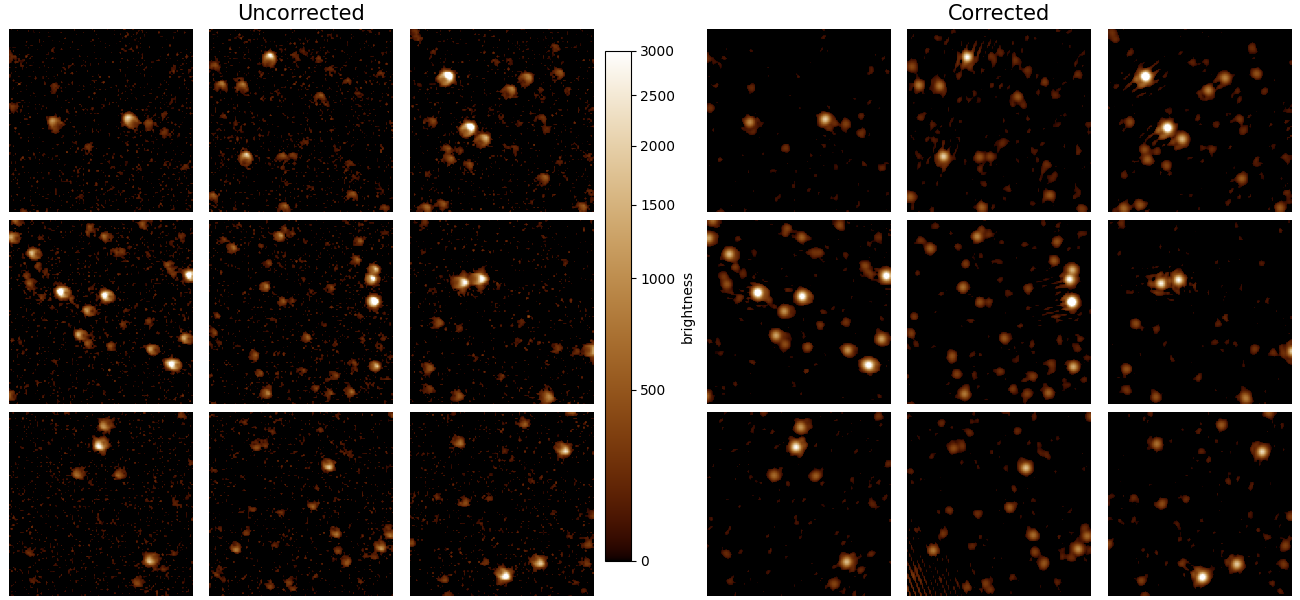}
    \caption{Image patches in a DASH image before and after PSF \new{homogenization}. Left: Stars in each region exhibit a varying PSF before correction. Right: After application of a transfer PSF, the stars are more uniform across the image.  The panels are gamma-corrected to highlight the periphery of the
    original and \new{homogenized} PSFs.}
    \label{fig:dash}
\end{figure*}

The Demonstration Airglow-Subtracted Heliospheric imager (DASH) was a prototype
instrument to demonstrate wide-field 
heliospheric imaging from near Earth.  DASH was deployed approximately 10 meters
from Earth, on a rooftop in the Colorado Rocky Mountains for 90 days in 2014. 
It was used to test
deep subtraction of multiple foregrounds/backgrounds including atmospheric 
effects and the starfield
\citep{DeForest_etal2015,DeForest_etal2015b}.  The instrument 
comprised an automated observatory housing, a corral-type baffle, a
commercial wide-angle lens, a commercial cooled-CCD camera, and
control electronics.  DASH collected 30 second cadence images nightly in the 
Western sky.  Data 
reduction involved very precisely tracking all visible stars as they crossed through the 
DASH FOV over
the course of each night's three-hour observing ``run''.

The Nikon wide-angle lens exhibited significant aberration at wide angles,
complicating analysis, as shown in Figure \ref{fig:dash}.  While the central
portion of the image is close to ideally sampled (with a PSF core FWHM of
approximately 2 pixels), the edges of the field of view exhibit some spherical
aberration, astigmatism, and broad sagittal rays; this meant that the shape
of a star changed considerably as it crossed the FOV.  In the 2015 analysis, the
DASH team blurred the starfield to a FWHM of 7.5 pixels to render these artifacts
irrelevant to the analysis.  We now demonstrate repair of the DASH
starfield to a uniform target Gaussian PSF with FWHM of just 3.5 pixels,
increasing effective resolution by a factor of 2 while maintaining 
PSF uniformity across the full field of view.

We used a single DASH image to build the PSF model. 
SEP is a Source Extraction and Photometry library written in Python by 
\cite{Barbary2016}. We used SEP to identify stars in the image. We broke the
image into $256 \times 256$ neighborhoods, nine of which are shown in Figure
\ref{fig:dash}, panel (a).  We stacked patches around all the stars
centered in each 
neighborhood, to produce a median patch PSF image. We  used this composite
patch image as a PSF model
to correct the image, with $\alpha=2, \epsilon=0.3$, and a symmetric Gaussian
target PSF
with 
FWHM of 3.5 pixels.  
Figure \ref{fig:dash}, panel (b) shows the result of the \new{PSF} regularization, which
greatly reduced the aberrations from the commercial lens.

\subsection{\label{sec:punch} PUNCH}
In June 2022, an engineering model (EM) of the PUNCH
Wide Field Imager (WFI) camera was taken to McDonald Observatory to capture
dark sky images for calibration and test. 540 of these images are used in 
this study. While these data suffer from high noise as
the EM WFI camera was operating at roughly 30C, far from the in-flight design
temperature of -60C, 
the ensemble data set was sufficient to reveal the variation of the PSF 
across the field of view.  

\begin{figure*}[h]
    \centering
    \includegraphics[width=6.5in]{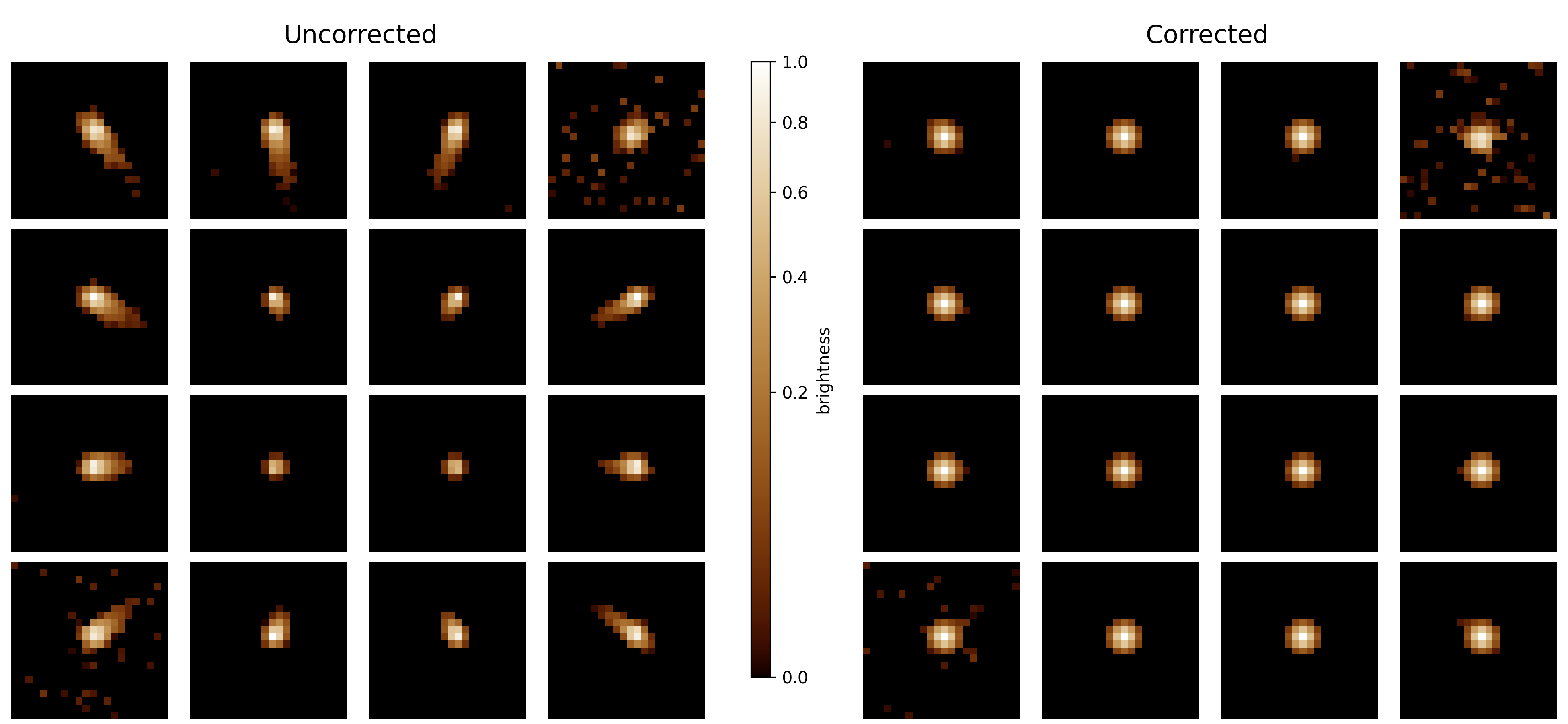}
    \caption{Average stars in PUNCH image regions before and after PSF \new{homogenization}. Left: Accumulated average star in each region (see text) exhibits a varying PSF before correction. Right: After application of a transfer PSF, the stars are uniform across the image and the PSF wings are reduced by a factor of order \new{30-100}.  The image is gamma-corrected to highlight the wings of each PSF.}
    \label{fig:punch}
\end{figure*}

We broke the WFI image into \new{$256\times256$}-pixel neighborhoods, and 
accumulated PSF images by 
co-aligning stars within each neighborhood of the
EM WFI image, as described in Section \ref{sec:accumulate-psf}, across \new{50} images in the dataset.  \new{Similar to Section \ref{sec:dash}, we utilized SEP to identify and extract stars.} 
Once we had extracted patches for all the selected stars imaged in each neighborhood,
we used the 
pixelwise median of all the patches to create an instrument PSF model for
that neighborhood. Figure 
\ref{fig:punch}, panel (a), shows a sampling of these PSF models for 
\new{sixteen} neighborhoods in the WFI FOV.

We then corrected the original images as described in Section 
\ref{sec:method}. We used
 a Gaussian target PSF with FWHM=2.25 
pixels, $\alpha=3$ and $\epsilon=0.3$.  Finally, we
re-assembled new PSF images from the corrected data.  Figure \ref{fig:punch},
panel (b), shows the PSFs after regularization.  The wings of the processed
PSFs are reduced in amplitude by a factor of order 30 compared to the original
data.

Some residual PSF effects are visible \new{in the peripheral neighborhoods} of Figure
\ref{fig:punch} (b); we attribute these to the high noise level in the source
images, and smaller number of samples of each star near the edges of the 
original field of view.

\section{\label{sec:discussion}Discussion} 

We have demonstrated \new{spliced linear PSF regularization to homogenize} variable telescope 
point-spread functions in astronomical applications, using direct
starfield imaging to recover and deconvolve for variable PSF 
across the field of view.  The method is useful for applications 
where uniform imaging over a wide field is more important than
the highest possible spatial resolution in a small portion of 
each image.  Further, it may improve future telescope design, by 
allowing designers to trade resources between telescopes
with
inexpensive optics (e.g., spherical mirrors in a Cassegrain 
telescope) and a large aperture,
vs. telescopes that minimize point-spread variability with more 
expensive optics (e.g., apheres in a Ritchey-Chretien telescope)
with a smaller aperture at the same price point.

We used a simple PSF \new{correction} method with batch processing 
of neighborhoods in a source image, each of which we
treated as having a single PSF susceptible to
direct linear inversion.  Direct linear treatment is equivalent
to construction of, and convolution by, a neighborhood-specific
``transfer PSF'' that converts the image in each neighborhood 
from its source PSF to a target PSF.  \new{The major advantage of this
method over more rigorous techniques is its efficiency: 2-3 orders 
of magnitude less computation is required than for existing
methods that consider a smoothly varying PSF.}

Direct linear known-PSF deconvolution has known limitations.  
In particular, it is equivalent to amplifying particular Fourier components
in the source image.  While deconvolution to sharpen an image beyond the 
core of its original PSF fraught with peril (because it involves strongly
amplifying many high-spatial-frequency components, along with any image
noise contained in them), deconvolution to remove PSF wings has far more 
modest effect on the noise level in the underlying data.  This is because PSFs
with broad wings and a sharp core do not strongly attenuate most image Fourier 
components \citep{DeForest_etal2009,Poduval_etal2013}; and because we are 
deconvolving not to an ideal one-pixel-wide PSF but to a target PSF with 
significant width, so that the final images are slightly oversampled by the
pixel grid.  Oversampling by 50\% or more is known to improve photometry by suppressing
artifacts related to feature placement relative to the pixel grid.  It also further band-limits the final
image to reduce amplification of high-spatial-frequency Fourier components.
In our examples, well under half of the
Fourier components of each image are multiplied by as much as a factor of 2, so
that overall image noise is nowhere increased by a factor greater than 1.5 and
generally much less.

Deconvolution in general and PSF regularization in particular work by 
comparative photometry across image pixels.  They are therefore limited by
the source images' photometric quality.  Calibration (flat-field) errors 
and/or detector nonlinearity induce errors that are amplified by the 
\new{PSF regularization} process.  

PSF regularization cannot significantly 
narrow the PSF core at any location in the source image -- so its best use
is to produce more uniform images, compromising optimal resolution at the
center of the image in favor of a uniform PSF across the entire field of view.
Potential applications include: (a) forming mosaics from multiple source
images, which can be \new{homogenized} for absolute uniformity of the resulting PSF
to reduce boundary effects in the final mosaic; (b) survey studies in which 
a large ensemble of features is detected either manually or automatically over
a wide FOV or even across multiple instruments with different imaging
characteristics, to eliminate PSF-based variations in the survey; (c) 
background removal problems such as the heliospheric imaging problem, in which
the steady starfield must be removed to reveal the much fainter dynamic
solar wind; (d) telescope optimization, in which a larger-aperture telescope
with spherical or other inexpensive optics may become preferable to a 
smaller telescope with more sophisticated (hence expensive) aspheric optics.

The technique we present here is \new{far from} the only \new{PSF correction} method that exists.
\citet{Schuler_etal2011} demonstrated PSF \new{homogenization} using second-order techniques, and \citet{Plowman_etal2022} have recently described a 
more general, but 
considerably more computationally intense, method of PSF
\new{homogenization}.  Plowman's method uses sparse matrix inversion techniques to
\new{homogenize} 
the PSF of the SPICE EUV spectral imager on 
board ESA's Solar Orbiter mission.  The present method is more specific to 
direct imaging in the particular (and most common) case of a slowly varying 
point-spread function that may be treated locally with linear methods; and is 
1.5-2.5 orders of magnitude less computationally intense than the more exact 
Schuler et al. or Plowman et al. methods.

\section{\label{sec:conclusion}Conclusions}

We have demonstrated a \new{direct} method of point-spread function \new{homogenization} 
that can equalize the PSF performance of a telescope or other optical system 
whose imaging characteristics vary slowly across the field of view, and that
is sufficiently robust and rapid to apply to large data sets.  
\new{Homogenized} images take on the appearance of having been collected by a single
ideal imager with uniform PSF, at modest cost in signal-to-noise
ratio (few-percent increase in noise level, for the examples we demonstrate here). 

The method will be deployed for the upcoming PUNCH mission, which benefits 
from the extreme uniformity in
modeling the starfield as a fixed background to be removed from that mission's
images.  PSF \new{homogenization} is useful for many astronomical and terrestrial applications
in which uniformity of imaging performance over a wide field is more important
than absolute maximal
resolution within a small subfield\new{, and a reference implementation of this 
rapid method is available for use as an open source package via github or Zenodo}

\begin{acknowledgments}
This work was funded through PUNCH, a NASA Small Explorer
mission, via NASA Contract No. 80GSFC18C0014.  
\end{acknowledgments}

\bibliography{PSF}{}
\bibliographystyle{aasjournal}

\section*{Appendix: Constructing the analytic model PSF}

In Section \ref{sec:model}, we constructed a synthetic instrument PSF from two Gaussians: a symmetrical Gaussian for the core brightness of the star and a second elliptical Gaussian to model aberration effects. Here, for completeness,
we reproduce the exact expressions used to produce the PSFs in Figure \ref{fig:model} in the main text. 

The general Gaussian function is just:
\begin{equation}\label{eqn:gauss}
g(x, y; a, b, c, x_c, y_c, h) =  h\,e^{-\left(a x'^2 + 2 b x'y' + c y'^2\right)}\,,
\end{equation}
where $(x_c, y_c)$ is the center of the Gaussian with height $h$, and $x'\equiv x-x_c$, $y'\equiv y-y_c$. The free parameters $a, b, c$ correspond to a positive definite matrix. More specifically,
\begin{align}
\label{eqn:gauss-params}
\theta &= \frac{1}{2} \arctan \left(\frac{2b}{a-c}\right), \theta \in [-45^\circ, 45^\circ] \\
\sigma_X^2 &= \frac{1}{2(a \cdot \cos^2 ]\theta + 2b \cdot \cos \theta \sin \theta + c \cdot \sin^2 \theta} \\
\sigma_Y^2 &= \frac{1}{2(a\cdot \sin^2 \theta - 2b \cdot \cos \theta \sin \theta + c \cdot \cos^2 \theta}
\end{align}

$\theta$ is the rotation angle of the elliptical Gaussian. $\sigma_X^2$ and $\sigma_Y^2$ are the variances. 

For our instrument PSF, we used
\begin{equation}\label{eqn:instrument-psf}
f(x, y) = g_{core}(x, y) + g_{tail}(x, y)
\end{equation}
Both the $g_{core}(x, y)$ and $g_{tail}(x, y)$ have a full complement of parameters as described in Table \ref{table:model-params}. The PSF windows are $32 \times 32$ pixels. We used $\alpha=3$ and $\epsilon=0.3$.   The results
are shown in Figure \ref{fig:model} in the main text.

\begin{table}[h!]
\centering
\begin{tabular}{||c c||} 
 \hline
 Parameter & Value \\ [0.5ex] 
 \hline\hline
 $h_{core}$ & 25 \\
 $x_{c, core}$ & 16 \\
 $y_{c, core}$ & 16 \\
 $\sigma_{x, core}$ & 1\\
 $\sigma_{y, core}$ & 1  \\
 $\theta_{core}$ & 0 \\ 
 $h_{tail}$ & 8 \\
 $x_{c, tail}$ & $1.5 \cos(\arctan((y-16)/(x-16))+45^\circ) + 16$ \\
 $y_{c, tail}$ &  $1.5 \sin(\arctan((y-16)/(x-16))+45^\circ) + 16$ \\
 $\sigma_{x, tail}$ & 5 \\
 $\sigma_{y, tail}$ & 1 \\
 $\theta_{tail}$ & $\arctan2(y, x)$\\ [1ex] \hline
\end{tabular}
\caption{Parameters for variable instrument PSF model.}
\label{table:model-params}
\end{table}

\end{document}